\documentclass{WileyChemistry-template}

\DeclareSIUnit\year{yr}

\usepackage[normalem]{ulem}
\usepackage[export]{adjustbox}

\title{\raggedright Prebiotic Vitamin B$_\text{3}$ Synthesis in Carbonaceous Planetesimals}

\author{
\begin{minipage}{\textwidth}
	Klaus Paschek,*\textsuperscript{+,[a]} Mijin Lee,\textsuperscript{+,[a]} Dmitry A.~Semenov,\textsuperscript{[a,b]} Thomas K.~Henning\textsuperscript{[a]}
\end{minipage}
}

\newcommand{\affiliation}{
\begin{itemize}

\item[{[a]}] K. Paschek*, M. Lee, Dr. D. A. Semenov, Prof. T. K. Henning\\
Max Planck Institute for Astronomy, Königstuhl 17, D-69117 Heidelberg, Germany\\
E-mail: paschek@mpia.de

\item[{[b]}] Dr. D. A. Semenov\\
Department of Chemistry, Ludwig Maximilian University of Munich, Butenandtstraße 5-13, House F, D-81377 Munich, Germany

\item[{[\texttt{+}]}] These authors contributed equally.
\end{itemize}
}

\renewcommand{\dedication}{
	\begin{minipage}{\textwidth}
	\end{minipage}
}

\renewcommand{\abstract}{Aqueous chemistry within carbonaceous planetesimals is promising for synthesizing prebiotic organic matter essential to all life. Meteorites derived from these planetesimals delivered these life building blocks to the early Earth, potentially facilitating the origins of life. Here, we studied the formation of vitamin B$_3$ as it is an important precursor of the coenzyme NAD(P)(H), which is essential for the metabolism of all life as we know it. We propose a new reaction mechanism based on known experiments in the literature that explains the synthesis of vitamin B$_3$. It combines the sugar precursors glyceraldehyde or dihydroxyacetone with the amino acids aspartic acid or asparagine in aqueous solution without oxygen or other oxidizing agents. We performed thermochemical equilibrium calculations to test the thermodynamic favorability. The predicted vitamin B$_3$ abundances resulting from this new pathway were compared with measured values in asteroids and meteorites. We conclude that competition for reactants and decomposition by hydrolysis are necessary to explain the prebiotic content of meteorites. In sum, our model fits well into the complex network of chemical pathways active in this environment.}

\newcommand{\keywords}{
    Meteorites \textbullet\ 
    Nitrogen heterocycles \textbullet\
    Origins of life \textbullet\ 
    Prebiotic chemistry \textbullet\ 
    Thermochemistry
}

\begin{document}

\emergencystretch 3em

\twocolumn[\vspace{-1.5cm}\maketitle\vspace{-1cm}
	\textit{\dedication}\vspace{0.4cm}]
\small{\begin{shaded}
		\noindent\abstract
	\end{shaded}
}

\begin{figure} [!b]
\begin{minipage}[t]{\columnwidth}{\rule{\columnwidth}{1pt}\footnotesize{\textsf{\affiliation}}}\end{minipage}
\end{figure}



\section*{Introduction}
\label{introduction}

The origins of life on Earth remain an intriguing and extensively studied topic across many disciplines, and the exact mechanism that sparked life is still unknown. However, recent studies have suggested that meteorites might have played a critical role in delivering an essential portion of prebiotic material to the early Earth.\cite{Chyba1992,Morbidelli2015,Pearce2017} Carbonaceous chondrites, carbon-rich meteorites, might have contributed a large fraction of the Hadean and Eoarchaean Earth's crust and upper mantle during the Late Veneer\cite{Varas-Reus2019,Fischer-Godde2020} (also referred to as the ``Late Heavy Bombardment'' or ``Late Accretion'') and delivered organic molecules to the various postulated environments that are thought to have provided the necessary conditions for the origins of life. Two prominent examples of these environments on Earth are the so-called ``warm little ponds'' located on the early Earth's first continental crust, and the ``hydrothermal vents'' located at the bottom of the primordial ocean.

The warm little pond hypothesis, proposed by Charles Darwin,\cite{Pereto2009} suggests that life originated in small bodies of water on the surface of the early Earth. These ponds would have provided the necessary conditions for life to emerge, such as a favorable temperature, shielding from the UV radiation by water, and a rich variety of organic molecules, possibly delivered there by meteorites.\cite{Pearce2017} By going through daily or seasonal wet-dry cycles, warm little ponds could have promoted the polymerization of these simple prebiotic molecules into highly complex organics, initiating molecular chemical evolution and possibly culminating in the origins of life.\cite{DaSilva2015,Becker2018,Becker2019,Damer2020}

Another favored site for the origins of life is hydrothermal vents (also called ``smokers''), where mineral-rich, superheated water emerges from the ocean floor. These unique environments are thought to have provided many of the necessary conditions for abiogenesis. The high temperature and mineral-rich waters at hydrothermal vents might have created the perfect conditions for organic molecules to form and evolve into more complex structures.\cite{Martin2008} The question whether the impacts could have made a significant contribution to the reservoir of prebiotic organics in these environments due to the high dilution in the vast ocean remains open. Recent evidence that the mass accreted via the impacts of carbonaceous asteroids during the Late Veneer could have been as high as $\sim\SI{0.3}{\percent}$ of the Earth's mass\cite{Varas-Reus2019,Fischer-Godde2020} makes it conceivable that the exogenous delivery of organics could have been important also for these deep water environments.

\begin{figure}[t]
    \centering
    \includegraphics[width=8.6cm]{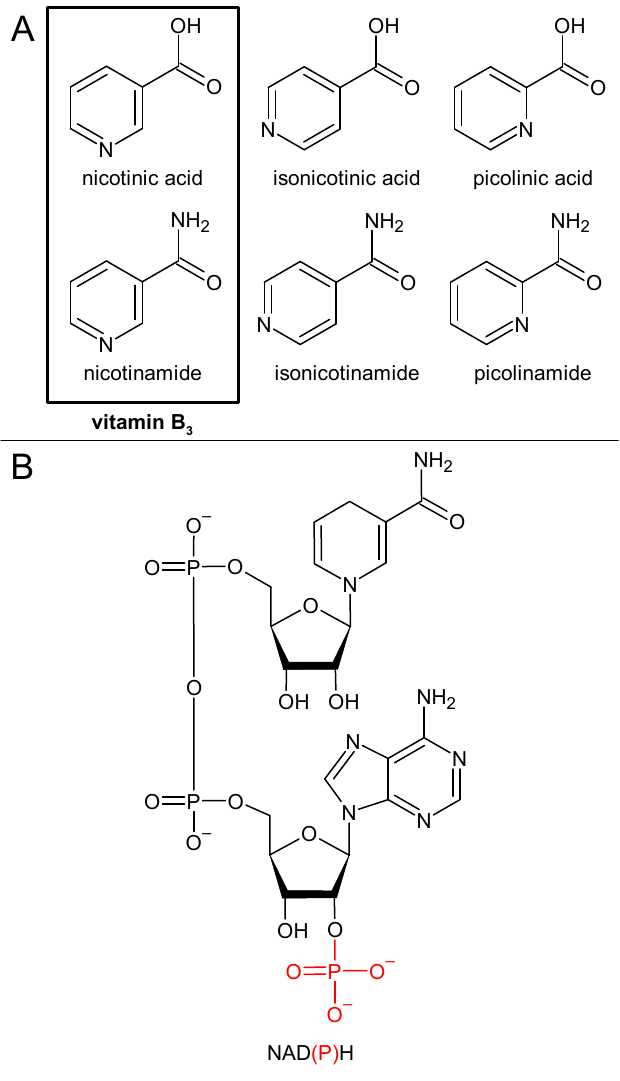}
    \caption{Structures of \textbf{A} vitamin B$_3$ (in box) and its isomers, and \textbf{B} the coenzyme nicotinamide adenine dinucleotide (phosphate), abbreviated as NAD(P)H, in its reduced form. The phosphorylated form is indicated in red.}
    \label{fig:structures}
\end{figure}

Among numerous key prebiotic organics such as fatty acids, amino acids, and nucleobases, the two forms of vitamin B$_3$ (rarely called vitamin PP), nicotinic acid and nicotinamide, also known as niacin and niacinamide, have been found in carbonaceous chondrites at concentrations of \SIrange{5}{715}{pbb}\cite{Smith2014,Oba2022} (see \cref{fig:structures}\textbf{A}). Furthermore, in the returned samples from the Hayabusa2 space mission that visited the near-Earth carbonaceous asteroid (162173) Ryugu, nicotinic acid was found in abundances of \SIrange{49}{99}{ppb}.\cite{Oba2023} This reinforces the idea that these organics have been already present in the parent bodies of carbonaceous chondrites in the early history of the solar system. The nicotinic acid and nicotinamide could either have been formed in their interior by aqueous chemistry, or they could have been inherited by the solar nebula from the interstellar medium and incorporated into the source material of the planetesimals.

Vitamin B$_3$ is essential to all life. It is used in a variety of biological functions and plays a critical role in cellular metabolism.\cite{Blum2015} It is the precursor of the coenzymes NAD$^+$\slash{}NADH and NADP$^+$\slash{}NADPH,\cite{Elvehjem1938} which are among the most important coenzymes in living systems (see \cref{fig:structures}\textbf{B}). In studies of autocatalytic reaction networks associated with the origin of metabolism, NAD has been identified as the key catalyst that might have kick-started redox chemistry at the time of the origins of life.\cite{Xavier2020} NAD stands for ``nicotinamide adenine dinucleotide'' and is, as the name implies, a dimer of two ribonucleotides. In the context of RNA nucleotides, nicotinamide, like adenine, takes the place otherwise occupied by nucleobases. This might link its first emergence to RNA molecules.

According to the so-called ``RNA world'' hypothesis, RNA might have been the starting point of chemical evolution, as some RNA molecules were shown to be able to store genetic information while catalyzing other chemical reactions, including their own polymerization, and replicating themselves.\cite{Kruger1982,Guerrier-Takada1983,Guerrier-Takada1984,Zaug1986,Cech1986,Johnston2001,Vaidya2012,Attwater2018,Cojocaru2021,Kristoffersen2022} These are also called ``ribozymes'' and are thought to have played a crucial role in the origins of life.

Replacing a canonical nucleobase in a ribonucleotide dimer with nicotinamide leads to NAD. It has been suggested that many coenzymes, including NAD, FAD, acetyl-CoA, and F420, comprise ribonucleotide moieties that may be surviving remnants of covalently bound coenzymes in an RNA world.\cite{Shen2011} Thus, nicotinamide and its role in NAD might be intimately linked to the origins of life in general, possibly as a critical link in an ``RNA-first-proteins-second world'' or an ``RNA-and-proteins-side-by-side world'' hypotheses.

The different forms of NAD are essential for various metabolic reactions in the cell throughout life. Examples include cellular redox reactions, energy production (synthesis of ATP, adenosine triphosphate),\cite{Friedkin1949,Rich2003} protection from oxidative stress,\cite{Rush1985,Couto2013} DNA repair,\cite{Wilkinson2001,Bürkle2005} cell signaling,\cite{Clapper1987,Ziegler2000,Berger2004,Guse2004} 5' modification of cellular RNA,\cite{Chen2009} among others.

All of these examples highlight the immense biological relevance of vitamin B$_3$ as a crucial precursor of NAD in the RNA world. The other necessary building blocks for NAD have also been found in meteorites,\cite{Stoks1981,Callahan2011,Pearce2015,Furukawa2019} and their synthesis mechanisms have already been explored in the aqueous environment of the interior of meteorite parent bodies.\cite{Pearce2016,Paschek2022,Paschek2023} The nucleobase adenine was found in carbonaceous chondrites\cite{Stoks1981,Callahan2011,Pearce2015} and its synthesis has been explained by previous models that agree with the measured values.\cite{Pearce2016,Paschek2023} The same is true for ribose, which was also found in carbonaceous chondrites,\cite{Furukawa2019} and its synthesis was investigated in a previous study.\cite{Paschek2022} Furthermore, the mineral schreibersite, which is capable of providing phosphate groups and phosphorylating nucleosides,\cite{Gull2015} is present in enstatite iron meteorites.\cite{Pasek2005,Bryant2006,Pirim2014} This makes nicotinamide the only piece needed for a complete building kit of NAD that could have been delivered to the early Earth by meteorites.

By pushing to understand its synthesis in space, this study aims to complete our knowledge of this meteoritic NAD building kit in order to better understand how meteorites might have contributed to the origins of life. All the ingredients could have been readily available in the early solar system and only needed to be assembled upon exogenous delivery onto the early Earth. With NAD as a subbranch in an emerging RNA world,\cite{Shen2011} vitamin B$_3$ in carbonaceous chondrites might have supported the origins of life itself. Moreover, by considering the role of meteorites in delivering essential organic matter to the warm little ponds and hydrothermal vents, as well as the potential role of vitamin B$_3$ in prebiotic chemistry, we can also better understand the conditions that are necessary for life to emerge, either on the early Earth or rocky exoplanets.

\subsection*{Carbonaceous chondrites}\label{sec:carbonaceous_chondrites}

Carbonaceous chondrites are widely recognized as the most interesting subclass of rocky meteorites in the context of prebiotic chemistry. They are undifferentiated and pristine bodies in character, close to their parent body material,\cite{Trigo-Rodriguez2019} and have early solar-like chemical compositions (minus the highly volatile elements).\cite{weisberg2006systematics} The name was coined due to their high carbon content $\sim\SI{5}{\percent}$ by weight, and they all have a preponderantly dark appearance.\cite{botta2002extraterrestrial,glavin2010effects,cobb2014nature} They contain up to \SI{20}{\percent} water and have an average porosity of typically $0.2$.\cite{Mason1963,Flynn1999,Macke2011}

Carbonaceous chondrites are divided into subgroups of distinctive composition. It is assumed that all members of a subgroup are descended from the same parent body or several very similar parent bodies formed in the same region in the early solar system.\cite{Bermingham2020} Some fragments resulting from mutual planetesimal collisions have been impacting the proto-Earth, carrying organics with them.\cite{Pearce2017} Some of these parent bodies might have remained asteroids in the solar system to this day.\cite{burbine2002meteoritic}

Each subgroup of meteorites is named after the first letter of the name of the first recovered fall. An example subgroup is CM (Mighei-like), which comprises carbonaceous chondrites that contain around \SI{9}{\percent} water by weight,\cite{rubin2007progressive} as well as aqueously altered minerals,\cite{palmer2011aqueous} with many classified as petrologic type 2. The distribution and associations of these hydrated phases provide insight into the role of water in the early solar system.\cite{palmer2011aqueous} CM chondrites are metal-deficient compared to other subgroups, such as CO (Ornans-like) chondrites, since metals are highly susceptible to aqueous alteration.\cite{kimura2011fe} Nevertheless, CM chondrites contain an array of metals in near-solar abundances, especially a significant amount of nickel.\cite{cobb2014nature} Palmer \& Lauretta\cite{palmer2011aqueous} and Kimura et al.\cite{kimura2011fe} discovered that CM chondrites contain an alloy of \ce{Fe}-\ce{Ni} and that kamacite grains are common in the Murchison and Murray meteorites.

CI (Ivuna-type) chondrites closely resemble the composition of the solar photosphere compared to other types of meteorites, and are thus generally accepted as the most pristine meteoritic material in the solar system.\cite{weisberg2006systematics,Anders1989} Cb-type asteroids, e.g., (162173) Ryugu, are considered to originate from the same parent body as CI chondrites, such as the Orgueil meteorite.\cite{Yada2022,Nakamura2022,Pilorget2022,Yokoyama2022,Oba2023}

\subsection*{Initial Concentrations of Reactants}\label{sec:initial_concs}

Understanding the initial molecular content and concentrations of reactive species inside carbonaceous chondrite' parent bodies is challenging because the original source material that formed these planetesimals no longer exists. Cobb et al.\cite{Cobb2015} and Pearce \& Pudritz\cite{Pearce2016} used abundances derived from spectroscopic observations of comets in their modeling studies aimed to explain the synthesis of amino acids and nucleobases. Comets are thought to have well-preserved reservoirs of volatiles that existed in the outer early solar system at $\sim$\SIrange{5}{20}{au} that are still accessible today.\cite{Rauer2008,Mumma2011} 

Nonetheless, comets are not representative of the reservoir of source material of carbonaceous chondrite parent bodies that formed in the solar system at $\sim$\SIrange{2}{3}{au}. For example, (13)~Egeria and (19)~Fortuna are regarded as potential CM chondrite parent bodies.\cite{Burbine1998} They are located close to the $3:1$ resonance with Jupiter at $\sim\SI{2.5}{au}$\cite{burbine2002meteoritic} and likely formed there. This was a warmer region of the solar nebula right outside the water snowline, as opposed to the outer cometary formation zone that was colder. 

This follows from state-of-the-art planet formation theories, explaining planet formation by accretion of pebbles, which migrate to the inner regions and form the first planetesimals due to streaming instabilities.\cite{Johansen2007,Ormel2010,johansen2014multifaceted,Klahr2020} These icy pebbles might originate from the formation region of comets and drift inwards closer to the proto-Sun. As they heat up, volatile species, e.g., \ce{CO} or \ce{O2}, start to desorb through the pores of the icy pebbles and are lost to space. Therefore, these volatiles are expected to be heavily depleted in the pebbles making up the source material of carbonaceous chondrite parent bodies. This means that one has to be careful when using the measured composition of comets\cite{Mumma2011} directly as a proxy for the initial composition of reactants involved in the prebiotic syntheses inside carbonaceous asteroids.

This depletion of volatiles was explored in a previous study.\cite{Paschek2023} The findings were based on experimental studies of mixtures of volatiles and water ice, as well as a review of the existing theoretical literature on the collapse of the solar nebula into the present solar system and the tracking of volatile ices. The predicted initial concentrations for reactants are listed in \cref{tab:concs}.

\setlength{\tabcolsep}{4.1pt}
\begin{table}
	\begin{center}
	\caption{Initial concentrations of reactants used in the model.\cite{Cobb2015,Goesmann2015,Paschek2023} Concentrations are normalized to water.}\label{tab:concs}
		\begin{tabular}{llc}	
\toprule		
Molecule & Name & Concentration \\
$i$ &  & [${\mathrm{mol}_i\cdot{}\mathrm{mol}_{\ce{H2O}}^{-1}}$] \\
\midrule
\ce{H2O} & water & 1 \\
\ce{NH3} & ammonia & 7$\times10^{-3}$ \\
\ce{HCN} & hydrogen cyanide & 2.5$\times10^{-6}$\\
\ce{H2CO} & formaldehyde & 6.6$\times10^{-4}$ \\
\ce{HOCH2CHO} & glycolaldehyde & (0.05-4)$\times10^{-4}$ \\
\ce{CH3CH2CHO} & propanal & (0.00125-1)$\times10^{-4}$\\
\bottomrule	
	\end{tabular}
	\end{center}
\end{table}

The abundances for volatile species (\ce{HCN} and formaldehyde) are adopted from our previous study,\cite{Paschek2023} while abundances for non-volatile reactants (\ce{NH3} and glycolaldehyde) necessary for the present study were adopted from another study\cite{Cobb2015} about amino acid synthesis in carbonaceous planetesimals, citing spectroscopic measurements in the comet {C/1995~O1~(Hale-Bopp)}.\cite{Mumma2011} The only newly added abundance is for propanal, also known as propionaldehyde. This aldehyde was detected in situ by the lander Philae,\cite{Goesmann2015} part of the ROSETTA mission, visiting and landing on comet 67P/Churyumov–Gerasimenko. It is important to note that Hale-Bopp and 67P belong to different classes of comets. Hale-Bopp originated in the Oort cloud very far outside the solar system, whereas 67P is part of the Jupiter family, orbiting closer to the Sun.\cite{Altwegg2019} Therefore, volatiles such as \ce{CO} or water should be more depleted in 67P compared to Hale-Bopp. More refractory ices, e.g., glycolaldehyde or propanal, should have remained more intact, but their abundance normalized to water might have changed, as water started to be lost to space. Therefore, we adjusted the initial concentrations of propanal by using the one for glycolaldehyde measured for Hale-Bopp\cite{Mumma2011} and using the ratio of $1/4$ between the two measured by the Philae lander on 67P\cite{Goesmann2015} (see \cref{tab:concs}).

\section*{Results and Discussion}
\label{results_discussion}

\subsection*{Reaction Pathway}

In their experiments, Cleaves \& Miller\cite{Cleaves2001} demonstrated the synthesis of nicotinic acid by combining aspartic acid with glyceraldehyde or dihydroxyacetone. This mixture of an amino acid and a sugar precursor is highly feasible to be present in carbonaceous planetesimals, as their respective synthesis mechanisms were shown to operate in these environments.\cite{Cobb2015,Paschek2022}

The favored mechanism for amino acid synthesis in planetesimals is the Strecker synthesis,\cite{Cobb2015} which involves aldehydes, \ce{HCN}, ammonia, and water (see previous Section on \nameref{sec:initial_concs} and \cref{tab:concs}). Likely resulting from this Strecker synthesis, aspartic acid was detected in a wide range of carbonaceous chondrites, and thus was also available for chemical reactions.\cite{cobb2014nature,Kaplan1963,Glavin2021}

The sugar precursors glyceraldehyde and dihydroxyacetone are easily formed via the formose reaction, which was shown to successfully describe the synthesis of ribose inside carbonaceous planetesimals.\cite{Paschek2022} Other necessary reactants, formaldehyde and glycolaldehyde, are also available (see \cref{tab:concs}). The formose reaction requires catalysts providing an alkaline environment, e.g., hydroxides or carbonates,\cite{Iqbal2012} which have also been found in carbonaceous chondrites.\cite{Barber1981} Therefore, all the necessary reactants have been available, and the reaction pathway leading to vitamin B$_3$ was operational in our environment of interest.

\begin{scheme*}[p]
\begin{center}
\includegraphics[width=17.4cm]{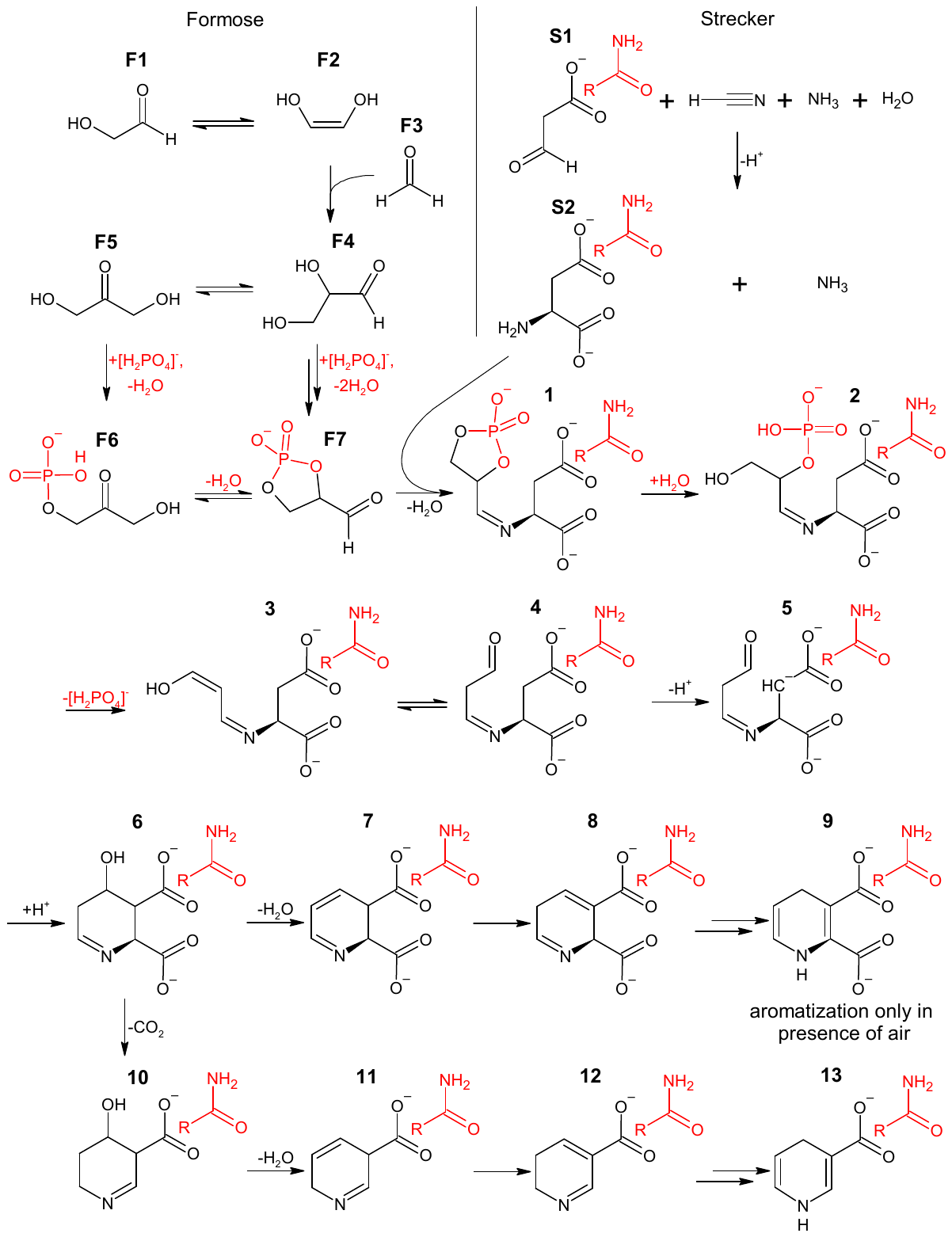}
\caption{Newly proposed reaction mechanism inspired by experimental studies by Cleaves \& Miller (2001).\cite{Cleaves2001} The phosphate and associated reaction steps are shown in red to indicate that phosphorylation increased the yield in the experiments; however, the synthesis of nicotinic acid was successful even without it, indicating it might be optional. Phosphates might not be readily available in meteorite parent bodies, but the reaction mechanism might still operate. The amide group serves as an alternative to the carboxyl group and is therefore also highlighted in red. Depending on the side chain of the starting aldehyde, either nicotinic acid or nicotinamide is formed in the Strecker synthesis.}
\label{sch:reaction}
\end{center}
\end{scheme*} 

We propose a new reaction pathway, shown in \cref{sch:reaction}, parts of which were developed in a private communication with O. Trapp (Department of Chemistry, Ludwig-Maximilians-University Munich, May 2023). It is a version of the proposal by Cleaves \& Miller\cite{Cleaves2001} adapted to the conditions inside carbonaceous planetesimals. In their experiments, they mainly formed quinolinic acid \textbf{9}, and nicotinic acid and nicotinamide in smaller yields \textbf{13}, depending on whether aspartic acid or asparagine \textbf{S2} were used, respectively. When starting with phosphorylated glyceraldehyde \textbf{F7} or dihydroxyacetone \textbf{F6}, the yield of these products increased significantly in comparison to starting with nonphosphorylated reactants. Therefore, we presume that phosphates act as a catalyst here.

The amino acids can be synthesized in the Strecker synthesis, starting from an aldehyde \textbf{S1}. Depending on which aldehyde the Strecker synthesis starts with, one or the other amino acid is formed. This is indicated in \cref{sch:reaction} with the alternative side chain colored in red, and R as a placeholder for the remaining molecule given in its full structure in black.

Cleaves \& Miller suggested that the phosphate is immediately split off in the first step from dihydroxyacetone \textbf{F6}, leading to methylglyoxal in their proposed reaction mechanism. On the other hand, they showed that dihydroxyacetone phosphate yields more nicotinic acid than methylglyoxal does, which might favor dihydroxyacetone phosphate as the more likely intermediate in the reaction. This motivated us to propose a new mechanism, in which the phosphate group is removed later after reacting with the amino acid and allows closing the N-heterocycle.

After the formation of the imine \textbf{1}, the dephosphorylation allows the removal of an oxygen atom of \textbf{2}. Further, this new reaction step prepares for the tautomerization of the enol \textbf{3} to the aldehyde group of \textbf{4}. This allows the deprotonated carbon of the enolate in \textbf{5} to attack the electrophilic carbon in the aldehyde group on the opposite side of the ring, initiating the closure of the N-heterocycle \textbf{6}. Cleaves \& Miller explain this step with the reaction of this enolate with an alkene group on the opposite side of the ring. Nevertheless, since the nucleophile enolate reacts better with the aldehyde group in \textbf{5}, our new proposal seems more feasible.

It should be noted that the experiments were also yielding products using nonphosphorylated reactants. It is an open question whether carbonaceous planetesimals contained enough minerals to allow this phosphorylation. Iron meteorites contain the phosphate-rich mineral schreibersite,\cite{Pasek2005,Bryant2006,Pirim2014} but carbonaceous chondrites do not. Whitlockite, a form of calcium phosphate, was detected in the carbonaceous chondrites Ningqiang\cite{Hsu2006} (anomalous CV condrite) and Yamato-82094\cite{Shibata1994} (CO3.5).\cite{pravdivtseva2007xe} This mineral might potentially be able to initiate phosphorylation in carbonaceous planetesimals. In \cref{sch:reaction}, the phosphate and involved reaction steps are highlighted in red to indicate that the phosphorylation might be optional.

Charge migration in \textbf{6} allows for its decarboxylation to \textbf{10} and branching off from the path \textbf{7}, \textbf{8} to quinolinic acid \textbf{9}. This allows for a direct aromatization of the N-heterocycle, pushing the synthesis directly toward nicotinic acid or nicotinamide \textbf{13}. After the elimination of \ce{H2O} from \textbf{6}/\textbf{10} to \textbf{7}/\textbf{11}, quinolinic acid likely stays non-aromatized in the absence of air. Cleaves \& Miller performed their experiment under vacuum conditions. Following our mechanistic proposal, this would mean that in their experiments they either detected this non-aromatized form of quinolinic acid \textbf{9} or they had air leaking in. They did not provide a mechanistic explanation of how the decarboxylation that leads to the nicotinic acid detected in the experiment occurs. However, they suggested that this decarboxylation might occur prior to aromatization because they tested whether quinolinic acid is stable under the conditions investigated. This observation is consistent with our proposal where the charge migration and decarboxylation directly lead to the aromatization.

The presence of \ce{O2} in meteorite parent bodies is unclear, as it was only found in 67P,\cite{Bieler2015} not representing a pristine comet, making the assessment of its depletion throughout the early solar system's evolution challenging. The high reactivity of free oxygen in the environment of planetesimals containing many potential reaction partners, e.g., metals,\cite{vanKooten2022} makes it hard to assess how long oxygen might remain available there. In carbonaceous chondrites, iron is mostly in its oxidized ferrous and ferric forms, with metallic iron in minor (often zero) amounts.\cite{Garenne2019} This might hint that oxygen is mostly bound in iron and other metal oxides. Further, we are unaware of any detection of \ce{HCO3}. Vitamin B$_3$ synthesis pathways inspired by industrial processes\cite{Lisicki2022} require either oxygen or \ce{HCO3} as reactants. Therefore, these pathways are unfeasible to explain synthesis in meteorite parent bodies. It also remains an open question if alternatives proposed to the Strecker synthesis involving oxygen\cite{Koga2022} are plausible in meteorite parent bodies. Dowler et al.\cite{Dowler1970} and Friedmann \& Miller\cite{Friedmann1971} demonstrated prebiotic pathways toward vitamin B$_3$ involving precursors formed by electrical discharges. We did not consider this mechanism further, as the reaction conditions seem unfeasible for the interior of meteorite parent bodies (electrical discharges seem unfeasible inside rocks), and to our knowledge, key reactants involved were not found in comets.

In the potential absence of free oxygen in carbonaceous planetesimals, the synthesis of vitamin B$_3$ in its aromatized form might even be favored over quinolinic acid (following \cref{sch:reaction}). The aromatic vitamin B$_3$ derivatives \textbf{10}-\textbf{13} have enhanced thermodynamic stability\cite{Chalk2019} compared to the non-aromatized quinolinic acid derivatives \textbf{6}-\textbf{9}, whose piperideine ring is highly reactive.\cite{Shvekhgeimer1998} Therefore, the final non-aromatized quinolinic acid derivative \textbf{9} might be less stable and more likely to decompose to other organics than vitamin B$_3$ \textbf{13}. Some planetesimals, which have the right balance of a big enough size $\gtrsim\SI{10}{\kilo\meter}$ and late enough time of formation $\gtrsim\SI{2.5}{\mega\year}$, can maintain aqueous conditions in their porous interiors for hundreds of thousands to millions of years.\cite{Paschek2023} When considering these long time scales, this might explain why quinolinic acid was indeed not found in carbonaceous chondrites. Smith et al.~(2014)\cite{Smith2014} explicitly searched for quinolinic acid in carbonaceous chondrites and did not detect it, while both forms of vitamin B$_3$ were found in the same study. The decarboxylation itself (\textbf{6}\ce{->}\textbf{10}) might be how to aromatize the N-heterocycle in the experiments by Cleaves \& Miller, leading automatically toward vitamin B$_3$ synthesis. This might confirm that our proposed reaction mechanism is appropriate to explain vitamin B$_3$ synthesis in carbonaceous planetesimals.

It is interesting to note that in the experiments by Cleaves \& Miller\cite{Cleaves2001} alkaline conditions strongly favored the vitamin B$_3$ synthesis, which is in agreement with the conditions necessary for the formose reaction,\cite{Iqbal2012,Barber1981} as mentioned above. Due to the hydroxides and carbonates found in carbonaceous chondrites,\cite{Barber1981} the aqueous interior of meteorite parent bodies might provide alkaline conditions favorable for the complete set of reactions in our proposed mechanism (see \cref{sch:reaction}).

In the Strecker synthesis, we are targeting the amino acids aspartic acid and asparagine (indicated as black and red structures \textbf{S2}, respectively). The requisite aldehydes in the reaction are 3-oxopropanoic acid and 3-oxopropanamide, respectively (black and red structure \textbf{S1}, respectively). As abundances for these specific aldehydes are not known in comets, we use propanal as a proxy for their abundances as the closest structural equivalent (see \cref{tab:concs}). The detection of a plethora of different amino acids in carbonaceous chondrites\cite{cobb2014nature} might implicate that the specific aldehydes might actually be present in the source material of carbonaceous chondrites and be available to the Strecker synthesis. It might be that the aldehydes were either not explicitly looked for or their abundances were below the detection limits of ex situ spectroscopic observations of comets.

In our newly proposed pathway (\cref{sch:reaction}), vitamin B$_3$ is formed in its reduced form \textbf{13}. This is the same form of nicotinamide as in the coenzyme NAD in its reduced state NADH (see \cref{fig:structures}\textbf{B}). The oxygen-poor environment of meteorite parent bodies might favor this pathway, and thus the prebiotic synthesis of vitamin B$_3$ in its reduced form, over the others mentioned above. It is a contentious debate how reduced or oxidized the early Earth's mantle and atmosphere were.\cite{Zahnle2020,Rollinson2017,Kuwahara2023,Wang2019,Nicklas2019,Kuwahara2023b,Trail2012,Armstrong2019,Matsui1986} It might be that carbonaceous chondrites falling to the early Earth provided a high reducing power. This might have been necessary for the emergence of this important coenzyme and the origins of life by powering metabolic reduction reactions in the perhaps otherwise oxidizing environment of the early Earth.

\subsection*{Simulated Vitamin B$_3$ Abundances}\label{sec:resulting_abundances}

\begin{figure*}[t]
\begin{center}
\includegraphics[width=17.4cm]{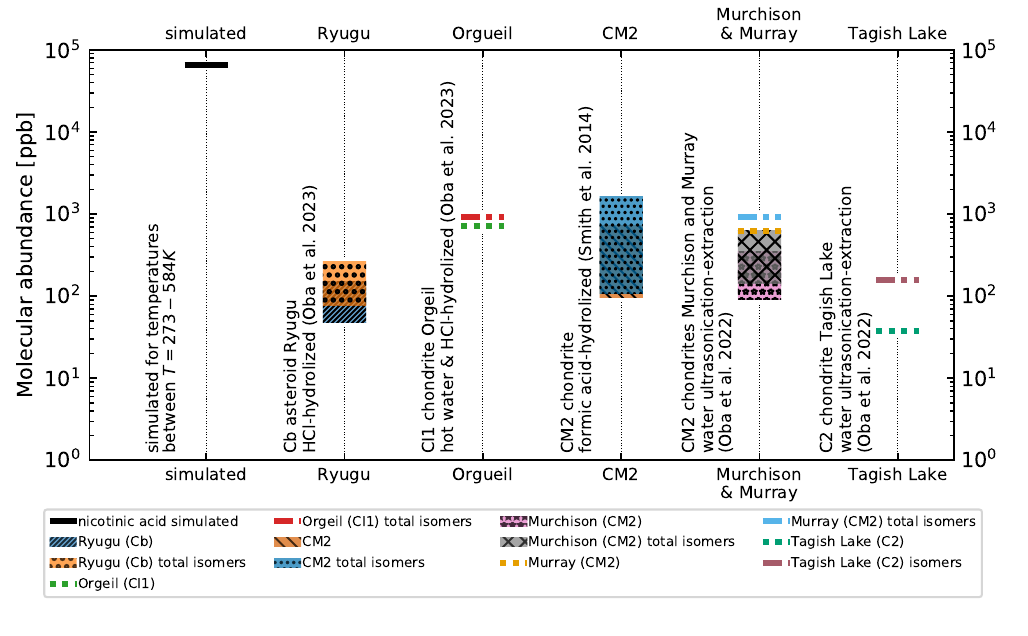}
\caption{Simulated nicotinic acid abundances compared to measured values in carbonaceous chondrites and asteroid (162173)~Ryugu. A rock density of \SI{3}{\gram\per\centi\meter\cubed}, a porosity of $0.2$,\cite{Mason1963,Flynn1999,Macke2011} and an ice density of \SI{0.917}{\gram\per\centi\meter\cubed} completely filling the pores (after all radionuclides have decayed and aqueous activity has ceased) were assumed as the properties of the planetesimal/carbonaceous chondrite hosting the chemical synthesis. The simulations were run at a pressure of \SI{100}{bar}.  From left to right, the plotted bars show the simulated molecular abundance of nicotinic acid (solid black line on the very left) for the whole temperature range of liquid water, as well as the abundances measured in samples of Cb asteroid (162173)~Ryugu collected during the Hayabusa2 spacecraft mission,\cite{Oba2023} in the CI chondrite Orgeil,\cite{Oba2023} several Antarctic CM2 chondrites,\cite{Smith2014} the CM2 chondrites Murchison and Murray, and the ungrouped C2 chondrite Tagish Lake\cite{Oba2022} as lines and shaded ranges described in the legend. Each time, the measured nicotinic acid abundance and the sum of all isomers (nicotinic acid, isonicotinic acid, picolinic acid) are given, since thermochemical equilibrium simulations cannot distinguish between isomers. The type of extraction method used (hot water, cold water ultrasonication, \ce{HCl}-hydrolyzed, formic acid) is denoted next to each panel. A tabulated version of the data presented here is available in the Supporting Information in Table S1.}
\label{fig:nac_sim}
\end{center}
\end{figure*} 

\begin{figure*}[t]
\begin{center}
\includegraphics[width=17.4cm]{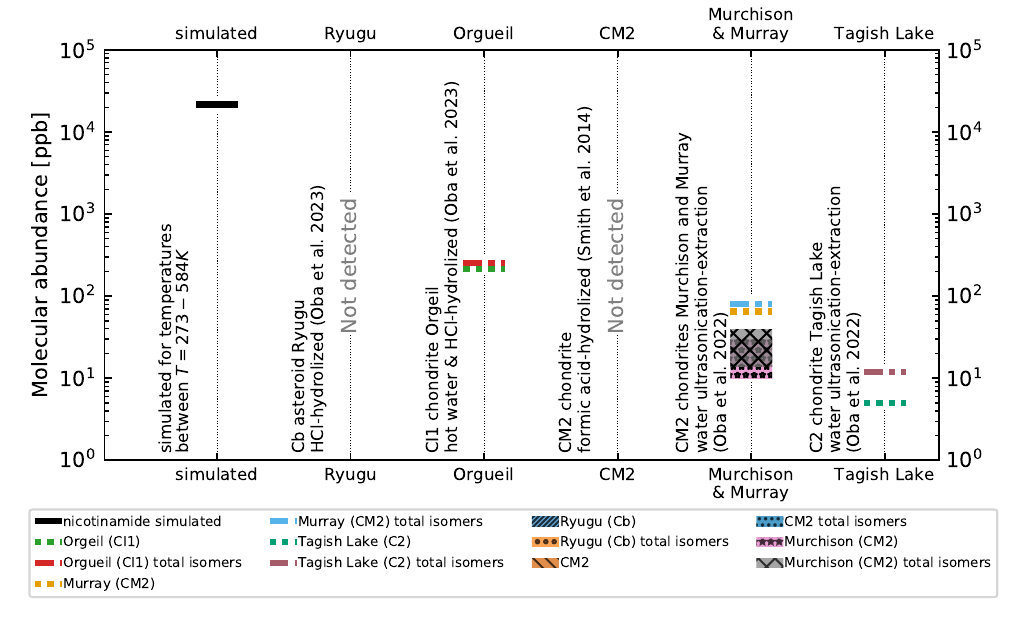}
\caption{Simulated nicotinamide abundances compared to measured values in carbonaceous chondrites and asteroid (162173)~Ryugu. A rock density of \SI{3}{\gram\per\centi\meter\cubed}, a porosity of $0.2$,\cite{Mason1963,Flynn1999,Macke2011} and an ice density of \SI{0.917}{\gram\per\centi\meter\cubed} completely filling the pores (after all radionuclides have decayed and aqueous activity has ceased) were assumed as the properties of the planetesimal/carbonaceous chondrite hosting the chemical synthesis. The simulations were run at a pressure of \SI{100}{bar}. From left to right, the plotted bars show the simulated molecular abundance of nicotinamide (solid black line on the very left) for the whole temperature range of liquid water, as well as the abundances measured in samples of Cb asteroid (162173)~Ryugu collected during the Hayabusa2 spacecraft mission,\cite{Oba2023} in the CI chondrite Orgeil,\cite{Oba2023} several Antarctic CM2 chondrites,\cite{Smith2014} the CM2 chondrites Murchison and Murray, and the ungrouped C2 chondrite Tagish Lake\cite{Oba2022} as lines and shaded ranges described in the legend. Each time, the measured nicotinamide abundance and the sum of all isomers (nicotinamide, isonicotinamide, picolinamide) are given, since thermochemical equilibrium simulations cannot distinguish between isomers. The type of extraction method used (hot water, cold water ultrasonication, \ce{HCl}-hydrolyzed, formic acid) is denoted next to each panel. If nicotinamide was not detected in the respective samples, this is noted in the respective panel. A tabulated version of the data presented here is available in the Supporting Information in Table S2.}
\label{fig:nam_sim}
\end{center}
\end{figure*} 

\subsubsection*{Testing the Thermodynamic Favorability}

To test if the newly proposed pathway in \cref{sch:reaction} is thermodynamically favorable, we performed thermochemical equilibrium calculations. The difference in Gibbs free energies of formation between products and reactants determines the favorability of a chemical reaction (see Computational Methods for more details). For that, an overall summarized reaction equation of \cref{sch:reaction} leads to 
\begin{equation}\label{eq:reaction_names}
\begin{aligned}
    &\ce{formaldehyde + glycolaldehyde}\\
    &\ce{+ 3-oxopropanoic\ acid / 3-oxopropanamide}\\
    &\ce{+ hydrogen\ cyanide + ammonia + water}\\
    &\ce{-> nicotinic\ acid / nicotinamide + carbon\ dioxide}\\
    &\ce{+ ammonia + water},
\end{aligned}
\end{equation}
or
\begin{equation}\label{eq:reaction_formula}
\begin{aligned}
    &\ce{H2CO + C2H4O2}\\
    &\ce{+ C3H4O3 / C3H5NO2}\\
    &\ce{+ HCN + NH3 + H2O}\\
    &\ce{-> C5H4NCOOH / C5H4NCONH2 + CO2}\\
    &\ce{+ NH3 + H2O},
\end{aligned}
\end{equation}
which we use in our thermochemical equilibrium calculations. Only if the difference between products and reactants is negative, does the reaction proceed spontaneously under the given environmental conditions inside meteorite parent bodies.

To test if the pathway is feasible, we formulate the following null hypothesis assuming the pathway is not able to explain the vitamin B$_3$ abundances measured in meteorites: if the vitamin B$_3$ abundances resulting from the calculations are zero or smaller than the measured ones, the pathway in \cref{sch:reaction} is unsuitable for the environment of meteorite parent bodies.

First, we need to define the physical conditions prevailing inside the parent bodies of carbonaceous chondrites. Thermodynamic simulations in a previous study\cite{Paschek2023} and hydrothermal simulations by Travis \& Schubert\cite{Travis2005} of carbonaceous chondrite parent bodies predicted the physical conditions allowing for liquid water in their porous interiors. The decay of radioactive isotopes, in particular, \ce{^{26}Al} as the dominant energy source in the early solar system, significantly heats up the planetesimal interiors. When assuming a pressure of \SI{100}{bar}, typical for a \SI{100}{\kilo\meter}-sized planetesimal,\cite{Pearce2016} the whole temperature range of liquid water of \SIrange{273}{584}{\kelvin} (\SIrange{0}{311}{\celsius}) can be reached inside the planetesimal. In a previous study,\cite{Paschek2023} we presented model planetesimals of radii between \SIrange{3}{150}{\kilo\meter} and times of formation after the formation of the solar system of \SIrange{0.5}{3.5}{Myr}. The bigger or the earlier formed a planetesimal is, the higher the reached temperatures can be. Bigger planetesimals can retain the energy generated by radioactive decay better. Earlier formed planetesimals contain a higher content of the short-lived \ce{^{26}Al} isotope dominating the early solar system.\cite{Cameron1977,Gaches2020}

In the previous study,\cite{Paschek2023} we explored the impact on the prebiotic chemistry of different temperature structures inside model planetesimals with different formation and evolution histories. We refrain from coupling the planetesimal models directly with the thermochemical equilibrium calculations in the present study, as we found no significant temperature dependence of the proposed reaction mechanism (see below). Nevertheless, higher temperatures used in the calculations presented here can be interpreted as representing either more central regions inside a planetesimal, or a generally bigger or earlier formed planetesimal at the same radial distance to its center.

Due to the high temperatures and pressure present in potential meteorite parent bodies, the pores in the whole central volume are filled with liquid water. This aqueous volume lies under a frozen layer at the surface of the planetesimal, shielding it from space. The water can remain liquid for several million to hundreds of millions of years.\cite{Paschek2023} This is consistent with simulations of equivalent model planetesimals by Travis \& Schubert\cite{Travis2005} and Lichtenberg et al.\cite{Lichtenberg2016} Rapid convection through the whole liquid volume allows for chemical equilibration and well-mixing of the entire system.\cite{Travis2005} Hence, thermochemical equilibrium calculations based on Gibbs free energies of formation as performed in the present study are a well-suited approach for this environment.

As already mentioned, the typical properties of carbonaceous chondrites and therefore likely also their parent bodies are a porosity of $\sim0.2$.\cite{Mason1963,Flynn1999,Macke2011} We assumed a general rock density of \SI{3}{\gram\per\centi\meter\cubed} and water ice density of \SI{0.917}{\gram\per\centi\meter\cubed} completely filling the rock pores of the frozen planetesimal (after the ceasing of radiogenic heating). This allows converting the molecular abundance resulting from the thermochemical equilibrium calculations into values comparable to measurements in real carbonaceous chondrites, usually given in parts per billion (\si{ppb}).

\cref{fig:nac_sim,fig:nam_sim} show the vitamin B$_3$ abundances resulting from the thermochemical equilibrium simulations. The leftmost black solid line shows the simulation result for the whole temperature range of liquid water. To compare this with the abundances found in carbonaceous chondrites, the bars and lines on the right show the measured abundances in different carbonaceous chondrites and the asteroid (162173)~Ryugu. The simulation results in abundances of \SI{6.64e4}{ppb} nicotinic acid and \SI{2.18e4}{ppb} nicotinamide, which are stable over the whole temperature range of liquid water at \SI{100}{bar}. These simulated abundances are around two orders of magnitude higher than the measured abundances in carbonaceous chondrites and asteroids.

As we do not see simulated abundances below the measured ones in carbonaceous chondrites, we conclude that the null hypothesis is rejected; the proposed pathway can not be ruled out and might explain how parts of the prebiotic vitamin B$_3$ synthesis in meteorite parent bodies proceeds. Only if the simulation had yielded lower abundances than measured, it would be completely unfit to explain the prevailing synthesis, as there would have to be another more dominating chemical process. We deduce that the pathway is well within the means of possibility.

\subsubsection*{Undetermined Initial Aldehyde Concentrations}

It is important to note that there is no information on the initial abundances for the reactants 3-oxopropanoic acid or 3-oxopropanamide (\textbf{S1} in \cref{sch:reaction}) measured in comets or predicted by solar nebula models. As already mentioned, we used the concentration of propanal measured in the comet 67P\cite{Goesmann2015} (see \cref{tab:concs}) as the aldehyde with the closest structural resemblance instead. Propanal might be the most general representative of this class of aldehydes with the lowest complexity. Thus, we would expect that this is most likely a strong overestimate of the actual concentrations of \textbf{S1} in the source material of carbonaceous chondrite parent bodies. Therefore, the resulting vitamin B$_3$ abundances presented here constitute the absolute upper limit possible.

\begin{figure*}[t]
\begin{center}
\includegraphics[width=17.4cm]{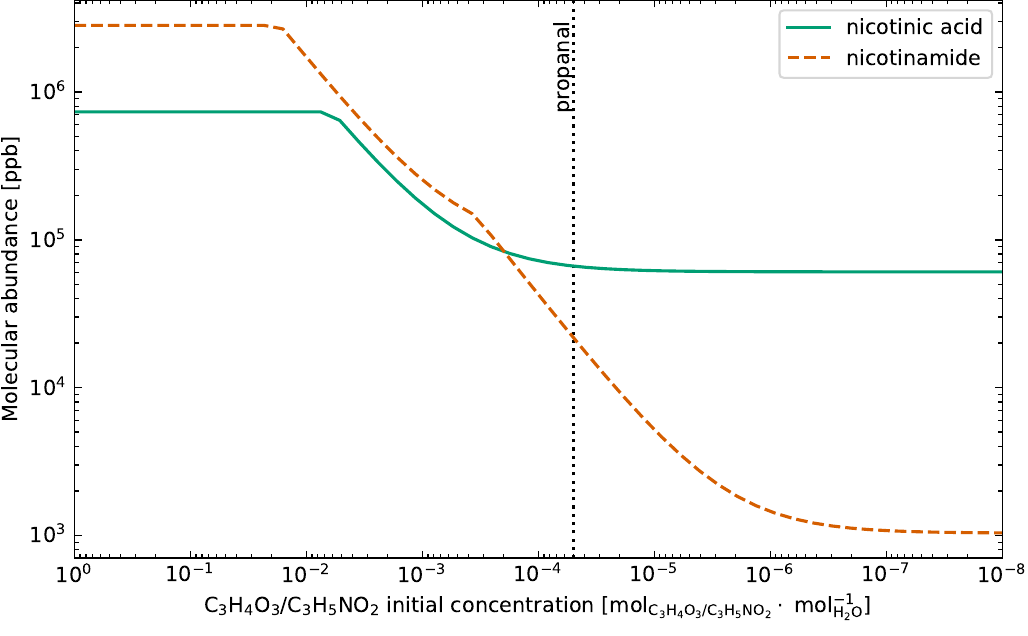}
\caption{Simulated vitamin B$_3$ abundances depending on variable initial concentrations of the aldehydes 3-oxopropanoic acid (\ce{C3H4O3}) and 3-oxopropanamide (\ce{C3H5NO2}), which are reactants in the Strecker synthesis (\textbf{S1} in \cref{sch:reaction}). All simulations were performed at \SI{0}{\celsius} and \SI{100}{bar}. The vertical dotted black line indicates the initial concentration of propanal (mean of the range given in \cref{tab:concs}), which was used in the simulations as a surrogate for these aldehydes that have not yet been detected in comets.}
\label{fig:variable_aldehyde}
\end{center}
\end{figure*}

To explore this further, we ran the same simulations for varying initial concentrations of 3-oxopropanoic acid and 3-oxopropanamide. The results are presented in \cref{fig:variable_aldehyde}. At very low initial aldehyde concentrations (on the right in \cref{fig:variable_aldehyde}), the resulting vitamin B$_3$ abundances are on a plateau because the thermodynamic favorability determines the balance of the reaction, not the initial reactant concentration. As the initial aldehyde concentration is increased, the resulting abundances begin to increase linearly. This corresponds to a regime where the increased supply of the aldehyde directly affects the balance of the reaction, leading to an increase in yield. It determines how much the synthesis yields and becomes completely depleted in the process. At even higher aldehyde concentrations, the resulting abundances level off at a saturation plateau because another reactant limits the synthesis by becoming exhausted. 

For nicotinic acid, assuming the cometary abundance of propanal does not significantly affect the resulting abundance, since a lower expected initial concentration of 3-oxopropanoic acid would change the resulting abundance in minor amounts moving further into the right plateau region (see \cref{fig:variable_aldehyde}). A lower initial 3-oxopropanamide concentration in comets has the potential to reduce the simulated nicotinamide abundance by about an order of magnitude. We conclude that using the cometary propanal abundance as a surrogate gives a good approximation for the achievable nicotinic acid abundance in this reaction mechanism, while for nicotinamide it is only an upper limit.

\subsubsection*{Comparison to Measured Abundances in Meteorites and Asteroid (162173) Ryugu}

One possible explanation for why the measured abundances are below the simulated ones (see \cref{fig:nac_sim,fig:nam_sim}) is that vitamin B$_3$ synthesis competes with the synthesis of other organic molecules drawing from the same initially available pool of reactants. As shown in previous studies, prebiotic synthesis of amino acids,\cite{Cobb2015} nucleobases,\cite{Pearce2016,Paschek2023} and sugars\cite{Paschek2022} draw from the same reactants in meteorite parent bodies. The autocatalytic formose reaction produces a plethora of complex sugars and decomposition products,\cite{Cleaves2015} competing for the glyceraldehyde \textbf{F4}/\textbf{F7} needed for vitamin B$_3$ formation.

The previous study of ribose synthesis\cite{Paschek2022} also did not consider the competition of the formose reaction for reactants with other pathways, and we were not yet aware of the pathway presented here. The result was that ribose was slightly overproduced in the simulations by factors of $\sim$\numrange{1.04}{100},\cite{Paschek2022} depending on if the minimum or maximum initial glycolaldehyde abundance in comets was assumed in the simulations (see also \cref{tab:concs} as this reactant also participates in the pathway of the present study). Knowing that the formose reaction is interlocked with vitamin B$_3$ synthesis via glycolaldehyde \textbf{F1} in \cref{sch:reaction} might explain why ribose was overproduced in these past simulations, as the difference between the simulated and measured ribose amounts in meteorites might actually account for the not yet known and not yet considered synthesis of vitamin B$_3$ operating in parallel. The formose reaction is much faster due to its autocatalytic nature in comparison to vitamin B$_3$ synthesis. The formose reaction completed in $\sim$\SIrange{20}{180}{\minute} in our past laboratory experiments\cite{Paschek2022} depending on the used catalyst and temperatures of $\le$\SI{60}{\celsius}. After this time, the maximum sugar abundances were reached and the decomposition of the products started to dominate. Cleaves \& Miller\cite{Cleaves2001} showed that vitamin B$_3$ synthesis was completed after around \SI{6}{\hour}, also at a temperature of \SI{60}{\celsius}. This might indicate that the autocatalytic formose reaction is much faster than vitamin B$_3$ synthesis, presumably resulting in most of the glycolaldehyde forming sugars, and only a minor fraction ending up as vitamin B$_3$ in the simultaneous synthesis. Nevertheless, both laboratory studies did not provide any kinetic parameters. Other factors, e.g., the initial concentrations of reactants or unknown rates of decomposition reactions play a role in the completion times of the reactions. Therefore, determining the reaction rates and comparing them directly would allow to test this hypothesis.

This might explain the overproduction in the present study (see \cref{fig:nac_sim,fig:nam_sim}), as this difference is actually forming an equivalent amount of sugars, including their decomposition products. There are many effective decomposition mechanisms active in the formose reaction, e.g., $\beta$-elimination, benzilic acid rearrangement, oxidation, etc.\cite{DeBruijn1986} Due to the presumably faster reaction rate of the formose reaction, the overproduction in the present study might be significantly larger in comparison to the previous investigations of ribose. Sugars and their decomposition products are major ``side products'' in vitamin B$_3$ synthesis, leading to the overproduction in the simulations, because competition with the formose reaction was not included in the model.

For a complete evaluation of the coupled formose reaction, its notoriously complex product mixture\cite{Cleaves2015} would need to be included. In the previous study,\cite{Paschek2022} we compensated for this by running the formose reaction in the laboratory and focusing only on ribose in relation to all formed pentoses. To fully understand the interlocking with vitamin B$_3$ synthesis, a complete analysis of all compounds forming in a combined formose and vitamin B$_3$ synthesis would need to be performed in experiments, which could then be used to tweak the simulations. This is beyond the scope of the present study. Our aim is to establish a first pathway of vitamin B$_3$ synthesis suitable for the environment of planetesimal interiors and to check its thermodynamic feasibility. In future studies, a simultaneous evaluation of the formose and vitamin B$_3$ synthesis might show if this can account for the overproduction seen in both the present and past models. If not, other yet unknown interlocked synthesis mechanisms producing other prebiotic molecules might exist. The newly proposed pathway in \cref{sch:reaction} is a first step in exploring the complex network of pathways governing the aqueous synthesis of the prebiotic organics found in carbonaceous chondrites.

Hexamethylenetetramine (HMT) was detected in carbonaceous meteorites and can be degraded to formaldehyde and ammonia upon hydrothermal treatment.\cite{Oba2020} HMT was shown to be formed by photochemical reactions and warming of interstellar ice analogs.\cite{Bernstein1995,MunozCaro2004,Vinogradoff2011} Since formaldehyde and ammonia are reactants in our proposed vitamin B$_3$ synthesis pathway (see \cref{sch:reaction} and \cref{eq:reaction_names,eq:reaction_formula}), HMT decomposing in the aqueous interiors of carbonaceous planetesimals could be another potential source of vitamin B$_3$ not included in the present model. If the presence of HMT in the interstellar medium is detected by observations, future studies might be able to estimate the initial concentration of HMT in the solar nebula and explore whether it has the potential to contribute significantly to vitamin B$_3$ synthesis in meteorite parent bodies.

Aspartic acid, which is one of the intermediate amino acids \textbf{S2} in \cref{sch:reaction}, was found in meteorites.\cite{Kaplan1963,Glavin2021} This indicates that a part of the available stock of amino acids participated in the vitamin B$_3$ synthesis and the rest remained in the parent body until after its phase of aqueous activity. Pearce et al.\cite{Pearce2016} showed that simulating the formation of several different prebiotic organics in thermochemical equilibrium models simultaneously often results in zero abundances or fails to reproduce the measurements in meteorites, as the model does not converge correctly due to its increased complexity. This limitation of the model means that we always obtain the upper possible limit using the whole pool of reactants for one specific pathway. The actual abundances are a weighted combination of the results for the synthesis of different organic molecules in parallel and partially yet unknown chemical mechanisms, drawing from overlapping sets of reactants, as mentioned above. This limitation of the thermochemical models would also be a challenge in future combined formose and vitamin B$_3$ models.

Not only the amount and allocation of the available reactants (which still needs to be explored further), but also processes involving the products might explain the overestimates in the simulations. In comparison to other vitamins, B$_3$ is relatively stable. On the other hand, B$_3$ is one of the least complex vitamins. Compared to the stability of other prebiotic molecules found in meteorites, e.g., nucleobases,\cite{Levy1998} vitamin B$_3$ (in this case nicotinamide) decomposes orders of magnitude faster,\cite{Yessaad2018} especially under hot ($\ge\SI{80}{\celsius}$) and alkaline conditions as presumed for carbonaceous parent body interiors, as mentioned earlier. This follows from comparing the pH-dependent decomposition rates and Arrhenius plots of nucleobases\cite{Levy1998} with the percentage of vitamin B$_3$ degradation at \SI{80}{\celsius} over \SI{60}{\minute},\cite{Yessaad2018} and might explain the overproduction in the present simulations. Potential destruction mechanisms of aqueous vitamin B$_3$ have not been included in the model but might reduce its final abundance found in meteoritic samples, especially over the long periods of radiogenic heating inside parent bodies.\cite{Paschek2023} It is difficult to quantify the exact pH and duration of aqueous activity in the parent body of a specific meteorite and therefore was not included in the model. Nevertheless, Smith et al.\cite{Smith2014} found a clear correlation between increasing aqueous alteration of the studied CM2 meteorites and a decreasing vitamin B$_3$ abundance, confirming that decomposition by hydrolysis plays a role in meteorite parent bodies.

Smith et al.\cite{Smith2014} also studied the hydrolysis of nicotinamide under the procedures employed to extract the organics from the meteorites. When using acid-hydrolyzed hot water (\SI{6}{M} \ce{HCl}, \SI{150}{\celsius}, \SI{3}{\hour}) for extraction, a common method also used by Oba et al.~(2023),\cite{Oba2023} purchased nicotinamide was \SI{100}{\percent} converted to nicotinic acid. This might also explain why nicotinamide could only be detected by the more gentle method by Oba et al.~(2022),\cite{Oba2022} using water extraction and ultra-sonication for \SI{10}{\minute} at room temperature without involving high temperatures or acids. In their other study Oba et al.~(2023),\cite{Oba2023} they could only find it in hot water extracts of the Orgueil CI meteorite at \SI{105}{\celsius} over \SI{20}{\hour}, which is at a lower temperature and over a shorter period in comparison to hot water extraction performed by Smith et al.\cite{Smith2014} In the same study Oba et al.~(2023),\cite{Oba2023} acid-hydrolyzed extraction (\SI{6}{M} \ce{HCl}, same as acid-hydrolyzed extraction by Smith et al.\cite{Smith2014}) of the same sample yielded no nicotinamide, but a slightly higher nicotinic acid abundance. This might indicate that in the acidic conditions of the extraction, nicotinamide was partially decomposed\cite{Yessaad2018} and partially converted to nicotinic acid.\cite{Smith2014} The method of extraction in meteorites itself seems to reduce the amount of organic material recovered, and hence might explain the discrepancy between the presented simulations and measurements.

In \Cref{fig:nac_sim,fig:nam_sim}, we also included the sum of all isomers of nicotinic acid (isonicotinic acid, picolinic acid) and nicotinamide (isonicotinamide, picolinamide) found in meteorites.\cite{Smith2014,Oba2022,Oba2023} This is due to the fact that thermochemical equilibrium simulations can not distinguish between isomers.\cite{Paschek2022,Paschek2023} Following from this limitation it might be that the simulated abundances actually represent the sum of all isomers.

In all the carbonaceous meteorite and asteroid extracts of Oba et al.,\cite{Oba2022,Oba2023} no picolinic acid was detected. Picolinamide was not detected in the Ryugu asteroid samples and the CI meteorite Orgueil, and in very minor amounts ($\le\SI{10}{ppb}$), which is similar to the background noise, in the CM2 meteorites Murchison and Murray as well as the ungrouped C2 meteorite Tagish Lake. Dominance of the nicotin- and isonicotin- isomers over picolin- isomers might be a direct consequence of the newly proposed mechanism in \cref{sch:reaction}. The decarboxylation by charge migration from \textbf{6} to \textbf{10} excludes that a carboxyl/amide group is at the 2-position of the pyridine ring of the final molecule. However, this is the case for the picolin- isomers or quinolinic acid mentioned above. On the other hand, starting with 4-oxobutanoic acid/4-oxobutanamide instead of \textbf{S1} results in glutamic acid/glutamine in the Strecker reaction. Combining this directly with glycolaldehyde \textbf{F1} (instead of glyceraldehyde \textbf{F4}/\textbf{F7}) to form an imine equivalent to \textbf{1} in \cref{sch:reaction} yields the isonicotin- isomers, following the same newly proposed mechanism. This is supported by the fact that glutamic acid was also found in meteorites.\cite{Kaplan1963,Glavin2021}

Nevertheless, Smith et al.\cite{Smith2014} found picolinic acid in their extracts from several CM2 meteorites, which is in direct contradiction to the missing detection in Murchison and Murray by Oba et al.~(2022),\cite{Oba2022} all members of the same meteorite group CM2 and potentially sharing the same parent body. They also performed proton-irradiation experiments of pyridine and \ce{CO2} ice mixtures. The underlying idea is that prebiotic organics formed in the interstellar and interplanetary medium on ice grains at low temperatures, and were incorporated into the source material of planetesimals. Meteorites might have inherited these organics and not (only) formed them in situ by aqueous chemistry, as presented here. Smith et al.\cite{Smith2014} claim to have found more picolin- than other isomers in these experiments, similar to the pyridine carboxylic acids detected in the extract of their most aqueously unaltered meteorite LEW 85311 studied. However, no respective chromatography-mass spectroscopy data were provided to support these findings. It is questionable if pyridine is readily available on ice grains, as this rather complex molecule was never detected in the interstellar medium.\cite{Charnley2005,Barnum2022,Heitkamper2022,Rap2023}

Anyway, there needs to be an explanation as to why Smith et al.\cite{Smith2014} found picolin- isomers in dominating amounts in meteorites, and Oba et al.\cite{Oba2022,Oba2023} did not. They might be inherited from ice grain processes and either very heterogeneously distributed in the same parent body, or the CM2 chondrites originate from several different bodies with different chemical histories. Another reason could be the contamination of the meteorite samples before their retrieval. At the same time, the asteroid material studied by Oba et al.~(2023)\cite{Oba2023} might have the highest chance of evading Earth's biosphere. Oba et al.~(2022)\cite{Oba2022} also refer to previous experiments on photon-irradiated \ce{H2O}, \ce{CO}, \ce{NH3}, and \ce{CH3OH} ice mixtures (Oba et al.~(2019)\cite{Oba2019}) to explain the synthesis of the N-heterocycles found in their meteoritic samples. The measured acid/amide ratios in the photochemical products were \numrange{0.8}{1.0}, and \numrange{9.3}{72.8} in the meteorite extracts.\cite{Oba2022} Oba et al.~explained the high ratios in meteorites with aqueous processing of photochemical products inherited from interstellar ice grains into the meteorite parent bodies. As an alternative to in situ aqueous chemistry, we recognize reactions on and inheritance from ice grains as another possible source of the vitamin B$_3$ in extraterrestrial carbonaceous material. More experimental work is needed here to find pathways feasible for vitamin B$_3$ synthesis in the interstellar medium. The same is true for aqueous pathways other than demonstrated by Cleaves \& Miller\cite{Cleaves2001} to study their interconnection with the mechanisms in \cref{sch:reaction}. This is necessary to better understand the origin of vitamin B$_3$ and its isomers in these extreme environments very different from the typical conditions studied in most chemical laboratories (extreme temperatures, no free oxygen available, etc.). It is an open question whether inheritance from ice grains or in situ aqueous synthesis is the primary source of vitamin B$_3$ and other prebiotic organics in meteorites.

\section*{Conclusion}\label{sec:conclusion}
	
Aqueous chemistry inside the warm porous rock of carbonaceous chondrite parent body planetesimals is a promising candidate for the prebiotic synthesis of vitamin B$_3$ in the early solar system. This could be a way to provide prebiotic molecules that are fundamental to all life as we know it and universal to all forming (exo)planetary systems. For the first time, we provide a detailed reaction mechanism suitable for this environment without any obvious caveats. Neither free oxygen nor electrical discharges are required, and the availability of necessary reactants is plausible. The origin of meteoritic organics and their contribution to the origins of life can only be understood by carefully assessing the physical setting and applying it to detailed chemical reaction mechanisms in an interdisciplinary approach.

We tested the thermodynamic favorability of our newly proposed reaction (see \cref{sch:reaction}) and evaluated the resulting abundances in comparison to measurements in carbonaceous chondrites and asteroids. We conclude that it fits well into the complex network of other prebiotic pathways, e.g., Strecker,\cite{cobb2014nature,Cobb2015} Fischer-Tropsch,\cite{Pearce2015,Pearce2016,Lai2019,Paschek2023} or formose\cite{Paschek2022} reactions active in the aqueous phase of planetesimals. More detailed modeling of competition between these pathways is needed, e.g., the interplay with the formose reaction. Additionally, decomposition reactions of vitamin B$_3$ either by hydrolysis in planetesimals or meteorite extraction in the laboratory need to be explored further to fully understand the abundances found by measurements in meteorites.

Inheritance of organics formed on the surface of ice grains by irradiation in the interstellar medium (often including reactions between radicals) is another promising synthesis pathway that needs to be acknowledged.\cite{Smith2014} After inheritance into meteoritic material, this might influence the final available organic reservoir brought to the early Earth by impacts, but this is beyond the scope of the present study.

It might be interesting to make 3-oxopropanoic acid, 3-oxopropanamide, 4-oxobutanoic acid, and 4-oxobutanamide new targets in surveys hunting for organics in comets. We hope the present work emphasizes the importance of these aldehydes for prebiotic synthesis in the early solar system and is able to stimulate cometary surveys to search for them despite their rather high complexity and specificity. If future surveys can detect these key reactants, this might bring us closer to understanding the prebiotic synthesis of the vital vitamin B$_3$ as well as proteinogenic amino acids in the interplanetary space of the arising solar system.

We hope the newly proposed mechanism in \cref{sch:reaction}, building on the previous work of Cleaves \& Miller,\cite{Cleaves2001} might encourage more research into prebiotic vitamin B$_3$. In the context of prebiotic synthesis, little attention seems to be given to it. This is despite the fact that it has the potential to connect the competing ``RNA first'' and ``metabolism first'' hypotheses in a unifying scenario, much like the central dogma of molecular biology in present life. 

\section*{Computational Methods}\label{sec:methods}

\begin{figure*}[t]
    \centering
    \includegraphics[width=17.4cm]{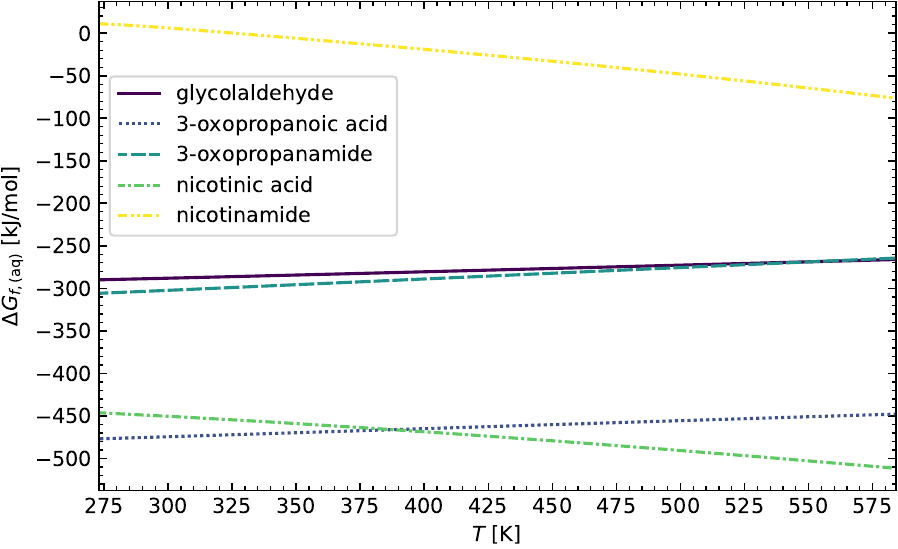}
    \caption{Gibbs free energies of formation $\Delta G_{f,\mathrm{(aq)}}$ as a function of temperature $T$ for the molecules not included in the CHNOSZ database. All energies are given in an aqueous solution at a pressure of \SI{100}{bar}, assuming an ideal infinite dilution.}
    \label{fig:Gibbs_add}
\end{figure*}

\subsection*{Thermochemical Calculations and Data}\label{sec:thermo_software_data}

The software ChemApp (version 740) provided by GTT Technologies was used to perform the thermochemical equilibrium calculations.\cite{Petersen2007} As inputs this software requires Gibbs free energies of formation for the molecules involved in the reaction (see also next Section on \nameref{sec:Gibbs_Free_Energy_of_Formation}). These were mostly obtained from the database CHNOSZ (version 1.3.6, 2020 March 16), providing a huge library of consistent Gibbs free energies of formation, obtained from a large set of experimental and theoretical studies.\cite{Dick2019}

The source code to set up and run the simulations, excluding the proprietary software ChemApp, is openly available on Zenodo\cite{klaus_paschek_2021_5774880} and as a Git repository (\url{https://github.com/klauspaschek/prebiotic_synthesis_planetesimal}). More details can be found in our previous study.\cite{Paschek2023}

\subsection*{Gibbs Free Energy of Formation}\label{sec:Gibbs_Free_Energy_of_Formation}

Each chemical molecule that takes part in a reaction as a reactant or product has a Gibbs free energy of formation $\Delta G_{f}$ that depends on temperature and pressure. The lower the value of $\Delta G_{f}$ the higher the probability that the molecule will form. The molecules should form spontaneously in the case of negative $\Delta G_{f}$ when the required reactants are available.

The $\Delta G_{f}$ as a function of temperature $T$ and pressure $p$ can be calculated by fitting the corresponding thermodynamic data $\Delta G_{f}$ from the CHNOSZ database to the function
\begin{equation} \label{eq:Gibbs_fit}
G_{f}(T, P) = a + bT +cT ln(T) + dT^{2} +eT^{3} + f/T + gP.
\end{equation}
The Gibbs coefficients $a$ through $g$ are necessary inputs for the equilibrium chemistry software ChemApp, which is used for the chemical reaction simulations.

As shown in previous studies,\cite{Cobb2015,Pearce2016,Paschek2023} in the temperature range of liquid water the Gibbs free energies are independent of pressure. This independence of pressure allows the pressure of the thermodynamic system to be set to a constant \SI{100}{bar}, making temperature and initial reactant concentrations the only dynamic simulation variables. This value is chosen as it is typically representing the conditions in porous planetesimals of up to hundreds of kilometers in radius due to lithospheric pressure.\cite{Pearce2016,Travis2005}

As the pressure dependence of $\Delta G_{f}$ is negligible, the Gibbs coefficient $g$ can be neglected.\cite{Pearce2016} Nevertheless, the boiling point of water remains the only pressure-dependent parameter of the model that needs to be considered. Our modeled reactions operate in the aqueous phase. It is assumed that when water evaporates, the prebiotic synthesis ceases, and lately formed organic molecules retain their present abundance. Each reaction must have a negative Gibbs free energy of reaction $\Delta G_{r}$ to be thermodynamically favorable, expressed as
\begin{equation} \label{eq:Gibbs_reaction}
\Delta G_{r} = \sum_{\mathrm{products}} \Delta G_{f} - \sum_{\mathrm{reactants}} \Delta G_{f}.
\end{equation}
Conversely, reactions with positive $\Delta G_{r}$ require activation energy to progress. This means that this would increase the total Gibbs free energy of the system $\Delta G$.

When the system reaches equilibrium, the chemical reactions cease, as at the current concentrations there is no longer a set of reactions that results in a negative $\Delta G_{r}$. This highlights the overall concept of thermodynamic chemical reaction simulation, which involves establishing initial concentrations of reactant molecules in the system and then calculating the resulting reactant and product concentrations that minimize $\Delta G$. The total Gibbs free energy of the system $\Delta G$ can be written as the combination of the Gibbs free energy of formation $\Delta G_{f}$ of each molecule
\begin{equation}\label{eq:Gibbs_total}
\Delta G = \sum_\mathrm{all} \Delta G_{f}.
\end{equation}
Catalysts have no role in minimizing Gibbs free energy computations because they do not commit molecules to the reaction. Catalysts only accelerate the reaction time by reducing the activation energy, which is a variable not used in equilibrium calculations.

\subsection*{Gibbs Free Energies of Formation for Species Missing in CHNOSZ}\label{sec:missing_Gibbs}

The CHNOSZ database contains only the Gibbs energy of formation for nicotinamide in its solid form $\Delta G_{f,\mathrm{(s)}}$. To obtain the Gibbs energy in an aqueous solution $\Delta G_{f,\mathrm{(aq)}}$, we used the formula
\begin{equation}\label{eq:Gibbs_solution}
    \Delta G_{f,\mathrm{(aq)}}(T) = \Delta G_{f,\mathrm{(s)}}(T) + R T \ln(s_\mathrm{sat}(T)),
\end{equation}
where $T$ is the temperature, $R$ is the ideal gas constant, and $s_\mathrm{sat}$ is the ideal saturated solubility in water in units of \unit{\mol\per\liter}, assuming infinite dilution and that Henry's law applies.\cite{Toure2016} We used measured temperature-dependent solubilities of nicotinamide in water\cite{Cysewski2021} and fit the data with a polynomial of second degree as a simple approximation of the temperature dependence. This allows us to extrapolate the measured data points for $s_\mathrm{sat}(T)$ over the whole considered temperature range. \cref{fig:Gibbs_add} shows the resulting temperature dependence of the aqueous Gibbs energies of formation, using \cref{eq:Gibbs_solution}.

Nicotinic acid is missing in CHNOSZ. We used measured literature values for the thermodynamic properties at standard conditions\cite{Knyazev2015,Di2017} and imported them into CHNOSZ. This allowed to calculate the Gibbs energies of formation over the whole considered temperature range.

Glycolaldehyde is also missing in the database, but we performed quantum chemistry calculations in a previous study\cite{Paschek2022} to obtain its Gibbs energies of formation. The same techniques were used to calculate the Gibbs energies of formation for the aldehydes 3-oxopropanoic acid and 3-oxopropanamide for the first time in a private communication with B.~K.~D.~Pearce (Department of Earth and Planetary Science, Johns Hopkins University, July 2023).

Using the software package Gaussian 09,\cite{g09} the atomic and molecular energies and entropies were determined to calculate the Gibbs free energy of formation. The Becke-3–Lee–Yang–Parr (B3LYP) hybrid density functional\cite{Stephens1994,Becke1993,Lee1988} and the polarizable continuum model (PCM) for aqueous solution effects\cite{Miertus1981,Cammi1995} were used, with geometry optimizations conducted using the 6-31G(d,p) basis set. Single-point energies and frequency were calculated using the 6-311++G(2df,2p) basis set. The Gibbs free energy of formation was determined using Ochterski's three-step method,\cite{Ochterski2000} incorporating enthalpy and entropy calculations at different temperatures. Some values were obtained from experiments\cite{Curtiss1997} and thermodynamic tables,\cite{Wagman1982} while an adjustment was made to the entropy of carbon (graphite) based on similar calculations for hydrogen and oxygen. A small error was introduced due to the use of gas-state carbon instead of carbon (graphite) for an enthalpy correction. More details can be found in the previous study\cite{Paschek2022} and the given references. \cref{fig:Gibbs_add} shows all the Gibbs energies of formation used in our simulation that are not available in the CHNOSZ database.

\section*{Acknowledgements}

The authors would like to thank Oliver Trapp for helping to understand the chemical reactions toward vitamin B$_3$ and deriving parts of the candidate prebiotic pathway. We would also like to thank Ben K.~D.~Pearce for performing quantum chemistry calculations to obtain Gibbs free energies of formation for some molecules missing in the CHNOSZ database. We thank Christelle Hiemstra for the stylistic review of the manuscript. K.P.\ acknowledges financial support by the Deutsche Forschungsgemeinschaft (DFG, German Research Foundation) under Germany's Excellence Strategy EXC 2181/1 - 390900948 (the Heidelberg STRUCTURES Excellence Cluster). K.P.\ is a fellow of the International Max Planck Research School for Astronomy and Cosmic Physics at the University of Heidelberg (IMPRS-HD). D.A.S.\ and T.K.H.\ acknowledge financial support by the European Research Council under the Horizon 2020 Framework Program via the ERC Advanced Grant Origins 83 24 28.

\section*{Conflict of Interest}

The authors declare no conflict of interest.

\begin{shaded}
\noindent\textsf{\textbf{Keywords:} \keywords} 
\end{shaded}


\setlength{\bibsep}{0.0cm}
\bibliographystyle{Wiley-chemistry}
\bibliography{main}

\clearpage


\section*{Entry for the Table of Contents}



\noindent\rule{13.4cm}{2pt}
\begin{minipage}{5.5cm}
\includegraphics[width=5.5cm]{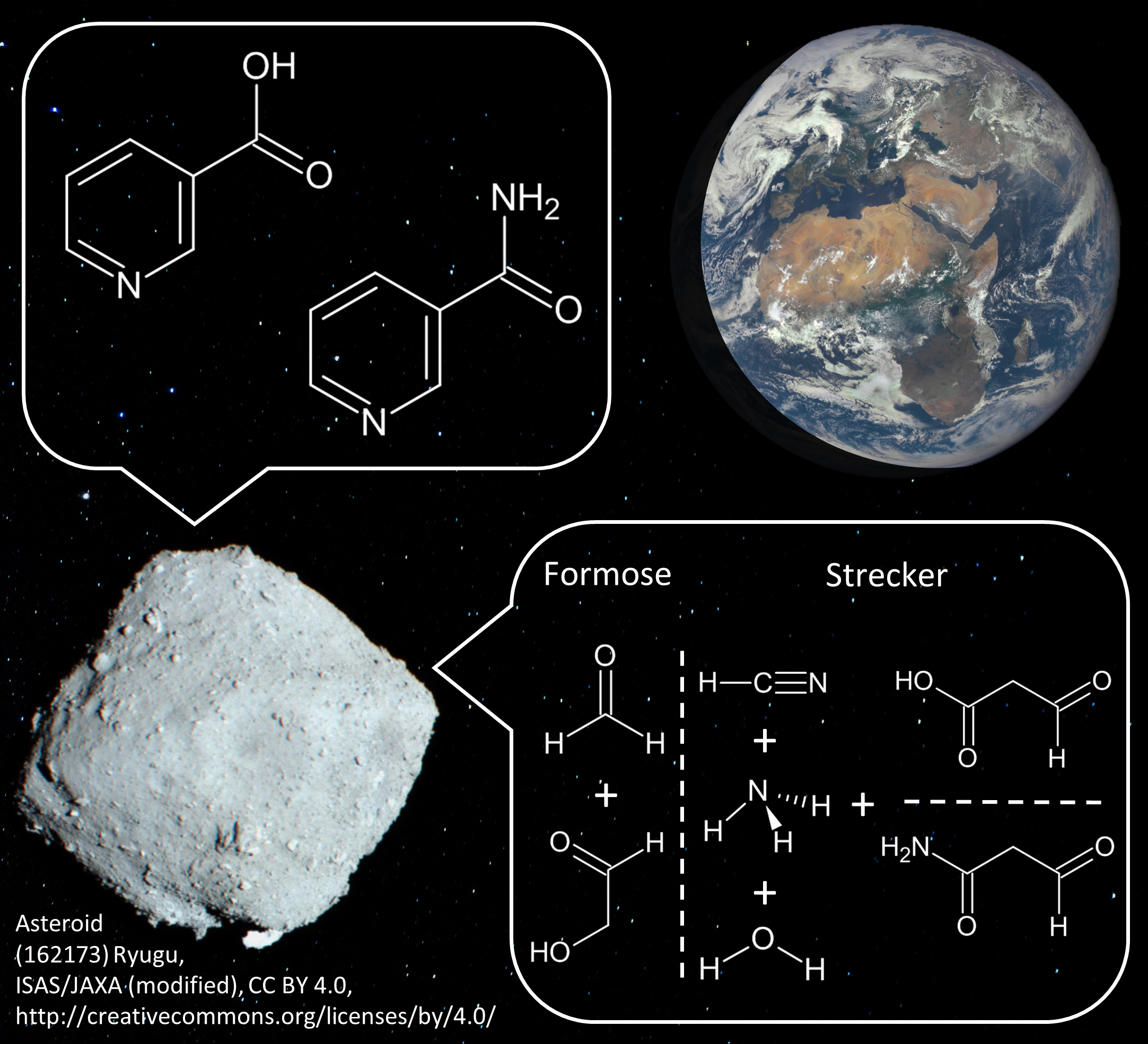} 
\end{minipage}
\begin{minipage}{7.8cm}
\large\textsf{Aqueous chemistry inside meteorite parent bodies allows the formation of prebiotic molecules crucial for all life in the early solar system. We present a reaction mechanism suitable for vitamin B3 synthesis in this environment, which is experimentally verified in the literature. Vitamin B3 was transported to the Hadean Earth by meteorite falls and initiated redox chemistry in the form of NAD(P)H, potentially participating in the origins of life.}
\end{minipage}
\noindent\rule{13.4cm}{2pt}


%
%

\end{document}